\documentstyle[12pt]{article}

\catcode`@=12
\newcommand{\be}{\begin{equation}}
\newcommand{\ee}{\end{equation}}
\newcommand{\bea}{\begin{eqnarray}}
\newcommand{\eea}{\end{eqnarray}}

\def\s12{\sin\theta_{12}}
\def\s23{\sin\theta_{23}}
\def\s13{\sin\theta_{13}}
\def\t12{\theta_{12}}
\def\t23{\theta_{23}}
\def\t13{\theta_{13}}

\def\l{\left}
\def\r{\right}
\def\dis{\displaystyle}
\def\la{\mathrel{\mathchoice {\vcenter{\offinterlineskip\halign{\hfil
$\displaystyle##$\hfil\cr<\cr\sim\cr}}}
{\vcenter{\offinterlineskip\halign{\hfil$\textstyle##$\hfil\cr<\cr\sim\cr}}}
{\vcenter{\offinterlineskip\halign{\hfil$\scriptstyle##$\hfil\cr<\cr\sim\cr}}}
{\vcenter{\offinterlineskip\halign{\hfil$\scriptscriptstyle##$\hfil\cr<\cr\sim
\cr}}}}}
\def\ga{\mathrel{\mathchoice {\vcenter{\offinterlineskip\halign{\hfil
$\displaystyle##$\hfil\cr>\cr\sim\cr}}}
{\vcenter{\offinterlineskip\halign{\hfil$\textstyle##$\hfil\cr>\cr\sim\cr}}}
{\vcenter{\offinterlineskip\halign{\hfil$\scriptstyle##$\hfil\cr>\cr\sim\cr}}}
{\vcenter{\offinterlineskip\halign{\hfil$\scriptscriptstyle##$\hfil\cr>\cr\sim
\cr}}}}}
\begin{document}
\thispagestyle{empty}
\vspace{0.5in}
\begin{center}
{\LARGE \bf 
Dark Matter candidate in a Heavy Higgs Model - Direct Detection Rates }
\vskip 0.1in
{\bf  Debasish Majumdar and Ambar Ghosal}
\vskip .5cm
Saha Institute of Nuclear Physics,\\
1/AF Bidhannagar, Kolkata (Calcutta) 700064, INDIA \\
\vskip .5cm
\end{center}
\begin{center}
\underbar{Abstract} \\
\end{center}

\noindent We investigate direct detection rates for Dark Matter candidates
arise in a $SU(2)_L\times U(1)_Y$ with an additional doublet Higgs
proposed by Barbieri, Hall and Rychkov. We refer this model as 
`Heavy Higgs Model'. 
The Standard Model Higgs mass comes out in this model very heavy 
adopting the few per cent
chance that there is no Higgs boson mass below 200 GeV.
The additional Higgs boson 
develops neither any VEV due to the choice of coefficient of the 
scalar potential 
of the model nor it has any coupling with fermions due to the 
incorporation of a discrete parity symmetry. Thus, the neutral 
components of the extra doublet are stable and can be considered 
as probable candidate of Cold Dark Matter.   
We have made 
calculations for three different types of Dark Matter experiments, 
namely, $^{76}$Ge (like GENIUS), DAMA (NaI) and XENON ($^{131}$Xe). Also 
demonstrated the annual variation of Dark Matter detection in case of all 
three detectors considered. 
\newpage
\noindent
A 'Heavy Higgs Model' has been proposed \cite{ref1} within the framework 
of $SU(2)_L\times U(1)_Y$ symmetry through the inclusion of another 
doublet and a discrete parity symmetry. This model is motivated to 
give an alternate way 
to control the quadratic divergence in the Higgs sector and 
to solve the naturalness problem. 
Again, results of electroweak precision tests are in
favour of light Standard Model (SM)
Higgs boson , $m_h < 186$ GeV at $95\%$ c.l \cite{ref01} 
with a central value
too low from the lower bound (114 GeV) obtained from direct search
experiments whereas, the present model is based on 
adopting the few per cent
chance that there is no Higgs boson mass below 200 GeV.
The basic feature of the model is as follows.

The additional doublet neither develops VEV nor it has 
any coupling with matter, however, the doublet reprsentation 
necessarily admits weak interaction and also scalar self interaction.
Due to absence of any coupling with matter of the extra doublet 
its neutral components are stable, and, therefore, those 
Lightest Stable Higgs (LSH) may be considered as probable 
candidate of cold dark matter. The zero VEV of the extra doublet 
is assured through the choice of coefficients of the scalar potential 
of the model and the discrete parity symmetry proposed in the model 
prohibits the extra doublet to couple with fermions.  

In the present work, we investigate Direct detection rates of those 
Dark Matter particles in view of ongoing and future Dark Matter experiments. 
\vskip 0.1in
\noindent
Let us first discuss relevant parts of the model. The model contains 
two doublet scalars $H_1$ and $H_2$. The $H_1$ scalar is the usual 
standard model doublet and the additional $H_2$ scalar 
is fermiophobic. The 
discrete parity symmetry imposed to achieve decoupling of $H_2$ field 
is given by  
\be
H_2\rightarrow -H_2
\ee
\noindent
while keeping all other fields 
invariant. The VEV of the Higgs fields are 
\be
H_1 = (0,v),\;\;\;\;\; H_2 = (0,0)
\ee
\noindent
and expanding around their minima we get 
\be
H_1 = \pmatrix{\phi^+\cr
               v + (h + i\chi)}
,\;\;\;\;\;
H_2 = \pmatrix{H^+\cr
               (S +iA)/\sqrt(2)}
\ee
\noindent
The scalar potential of the model is given by 
\begin{eqnarray}
V & = & \mu_1^2|H_1|^2 + \mu_2^2|H_2|^2 + \lambda_1|H_1|^4 + 
\lambda_1|H_2|^4 + \lambda_3|H_1|^2 |H_2|^2\nonumber\\
&&+ \lambda_4|H_1^\dagger H_2|^2 + \frac{\lambda_5}{2}
\left[(H_1^\dagger H_2)^2
+ h.c\right].
\end{eqnarray}
The mass of the usual Higgs boson $h$ in the present model is 
set around $m_h \sim 400 - 600$ GeV and the three  additional 
scalars 
as 
\begin{eqnarray}
m_I^2& = &\mu_2^2 + \lambda_I v^2 , I = {H,S,A}\nonumber\\
\lambda_H & = & \lambda_3\nonumber\\
\lambda_S & = &\lambda_3 + \lambda_4 + \lambda_5\nonumber\\
\lambda_A & = &\lambda_3 + \lambda_4 - \lambda_5
\end{eqnarray}
\noindent
and from the minimization of the scalar potential one obtains conditions for 
which the potential is stable, as 
\be
\lambda_{1,2} > 0\;\; {\rm and }\;\;
 \lambda_3,\;\lambda_L\;> 
-2{(\lambda_1\lambda_2)}^{1/2}
\ee
{\rm where}
\be 
\lambda_L = \lambda_3 + \lambda_4 - |\lambda_5|
\ee
The Higgs fields $S$, $A$ are stable as long as the parity symmetry (1) is 
unbroken and those stable neutral scalars appear to be a candidate for Cold 
Dark Matter in the universe. 
In order to see the prospect of direct detection 
of such a Dark Matter candidate, we 
consider  two processes of scalar-nucleon interaction  
i) via $Z$ boson exchange and ii) via Higgs scalar $h$ exchange. 

The amplitude comes out due to
$Z$ boson exchange is too large 
 (about 8-9 orders)
compared to the
existing CDMS collaboration experimental limits
\cite{cdms} on  spin-independent wimp-nucleon interactions 
from the two-tower run of the cryogenic dark matter search. Hence,
to neglect this contribution, a mass splitting between the two scalars
$S$ and $A$ are considered which is greater than the kinetic energy
of the dark matter in our galactic halo, so that the process due
to $Z$ exchange is kinematically forbidden.
We consider the other process due to $h$ exchange at the tree 
level for our analysis in the present work. 

In Ref. \cite{ref1}, the possible range of mass, $m_L$ for the LSH, 
to account for the Dark Matter content of the universe, is discussed.
For $m_L \ga m_W$ (mass of $W$ boson), the Dark Matter 
density $\Omega_{\rm DM}$ falls 
much below the acceptable value due to dark matter annihillation 
to $W$ bosons ($LL\longrightarrow WW$, where $L$ is generic representation 
of $S$ and $A$ fields). Therefore, massive LSHs can 
contribute to only a fraction of Dark Matter content. While LSH mass 
$m_L \la m_W$, the annihillation to $W$ bosons are kinematically 
disfavoured  and 
in particular $m_L \sim 60 - 70$ GeV can account for the Dark Matter
content of the universe with $\Delta m$, the mass difference between 
the LSH and NLSH (next lightest stable Higgs), $\sim 8$ GeV.
Therefore, we consider three LSH masses,
in the range 60 - 70 GeV in the present work. We also consider,
as an example, a value of $m_L$ in the $m_L >> m_W$ regime, which 
as mentioned before, can only account for just a fraction
of the total Dark Matter.

\vskip 0.1in
\noindent
The scalar-nucleon cross-section due to the later case comes out 
as 
\be 
\sigma(LN\rightarrow LN) = {\frac{m_r^2}{4\pi}}\left({\frac{\lambda_L}
{m_L m_h^2}}\right)^2 f_N^2 m_N^2
\ee
\noindent
where  $m_r$ is 
the reduced nucleon mass, $\lambda_L$ is the combination of scalar 
couplings defined in eqn.(4), $m_h$ is the mass of the 
Standard Model like Higgs scalar  and $f_N$ is the 
usual nuclear matrix element given by 
\be 
<N|\Sigma m_q q\bar q |N> = f_N m_N <N|N> 
\ee
and for the present analysis we set $f_N\sim 0.3$.

\vskip 0.1in
\noindent
In the present work, we calculate the possible direct detection rates
for such Lightest Stable Particles (or LSH) Dark Matter candidates 
discussed above, in the experiments 
like GENIUS (target material $^{76}$Ge) \cite{genius1,genius2}, 
DAMA (target material NaI) \cite{dama1,dama2,dama3} and XENON 
(target material $^{131}$Xe) \cite{xenon1,xenon2}. 

The direct detection of Dark Matter with a terrestrial detector uses the 
elastic scattering of Dark Matter candidate off the detector nuclei. 
As this cross-section is very small, the energy deposited by a Dark Matter
candidate of mass in the range 1 GeV to 1 TeV on a detector nucleus 
is not generally more than 
100 keV. Hence to perform this difficult task of Dark Matter detection 
a very low threshold detector condition is required.

Differential detection rate of Dark Matter per unit detector mass can be
written as
\begin{equation}
\frac {dR} {d|{\bf q}|^2} = N_T \Phi \frac {d\sigma} {d|{\bf q}|^2} \int f(v) dv\end{equation}
where $N_T$ denotes the number of target nuclei per unit mass of the detector,
$\Phi$ - the Dark Matter flux, $v$ - the Dark Matter velocity in the 
reference frame of earth with $f(v)$ - its distribution. The integration
is over all possible kinematic configurations in the scattering process.
In the above, $|\bf q|$ is the momentum transferred to the nucleus in
Dark Matter-nucleus scattering. Nuclear recoil energy $E_R$ is
                                                                                
\begin{eqnarray}
E_R &=& |{\bf q}|^2/2m_{\rm nuc} \nonumber \\
    &=& m^2_{\rm red} v^2 (1 - \cos\theta)/m_{\rm nuc}  \\
m_{\rm red} &=& \frac {m_L m_{\rm nuc}} {m_L + m_{\rm nuc}}
\end{eqnarray}
where $\theta$ is the scattering angle in Dark Matter-nucleus centre of 
momentum frame, $m_{\rm nuc}$ is the nuclear mass and $m_L$ is the 
mass of the Dark Matter.
                                                                                
Now expressing $\Phi$ in terms of local Dark Matter density $\rho_\chi$,  
velocity $v$ and mass $m_L$ and writing $|{\bf q}|^2$ in terms
of nuclear recoil energy $E_R$ with noting that $N_T = 1/m_{\rm nuc}$,
Eq. (10) takes the form
                                                                                
\begin{eqnarray}
\frac {dR} {dE_R} &=& 2 \frac {\rho_\chi} {m_L} \frac {d\sigma}
{d |{\bf q}|^2} \int_{v_{min}}^\infty v f(v) dv, \nonumber \\
v_{\rm min} &=& \left [ \frac {m_{\rm nuc} E_R} {2m^2_{\rm red}} \right ]^{1/2}
\end{eqnarray}
                                                                                
Following Ref. \cite{jungman} the Dark Matter-nucleus differential 
cross-section for the scalar interaction can be written as
\be
\frac {d\sigma} {d |{\bf q}|^2} = \frac {\sigma_{\rm scalar}}
{4 m_{\rm red}^2 v^2} F^2 (E_R) \,\,\, .
\ee
In the above $\sigma_{\rm scalar}$ is Dark Matter-nucleus scalar cross-section
and $F(E_R)$ is nuclear form factor given by \cite{helm,engel}
\begin{eqnarray}
F(E_R) &=& \left [ \frac {3 j_1(qR_1)} {q R_1} \right ] {\rm exp} \l ( \frac {q^2s^2}
{2} \r ) \\
R_1 &=& (r^2 - 5s^2)^{1/2} \nonumber \\
r &=& 1.2 A^{1/3} \nonumber
\end{eqnarray}
where thickness parameter of the nuclear surface is given by $s \simeq 1$ fm,
$A$ is the mass number of the nucleus and $j_1(qR_1)$ is the spherical
Bessel function of index 1.
                                                                                
The distribution $f(v_{\rm gal})$ of Dark Matter velocity $v_{\rm gal}$
with respect to
Galactic rest frame, is considered to be of Maxwellian form. The
velocity $v$ (and $f(v)$) with respect to earth rest frame can then be obtained
by making the transformation
\be
{\bf v} = {\bf v}_{\rm gal} - {\bf v}_\oplus
\ee
where $v_\oplus$ is the velocity of earth with respect to Galactic rest
frame and is given by
\bea
v_\oplus &=& v_\odot + v_{\rm orb} \cos\gamma \cos \l (\frac {2\pi (t - t_0)}
{T} \r )
\eea
In Eq. (17), $T = 1$ year, the time period of earth motion around the sun,
$t_0 = 2^{\rm nd}$ June, $v_{\rm orb}$ is earth orbital speed and
$\gamma \simeq 60^o$ is the angle subtended by earth orbital
plane at Galactic plane. The speed of solar system $v_\odot$ in the
Galactic rest frame is given by,
\bea
v_\odot &=& v_0 + v_{\rm pec}
\eea
where $v_0$ is the circular velocity of the Local System at the position of
Solar System and $v_{\rm pec}$ is speed of Solar System with respect to
the Local System. The latter is also called peculiar velocity and its value
is 12 km/sec. Although the physical range of $v_0$ is given by
\cite{pec1,pec2}
$170\,\, {\rm km/sec} \leq v_0 \leq 270$ km/sec (90 \% C.L.), in the present
work we consider the central value of $v_0 = 220$ km/sec. 
Eq. (17) gives rise to annual modulation of Dark Matter signal reported
by DAMA/NaI experiment \cite{dama1,dama2,dama3}.
                                                                                
Defining a dimensionless quantity $T(E_R)$ as,
\be
T(E_R) = \frac {\sqrt {\pi}} {2} v_0 \int_{v_{\rm min}}^\infty \frac {f(v)}
{v} dv\,\,
\ee
and noting that $T(E_R)$ can be expressed as \cite{jungman}
                                                                                
\be
T(E_R) = \frac {\sqrt {\pi}} {4v_\oplus} v_0 \l [ {\rm erf} \l ( \frac
{v_{\rm min} + v_\oplus} {v_0} \r ) -  {\rm erf} \l ( \frac
{v_{\rm min} - v_\oplus} {v_0} \r ) \r ]
\ee
we obtain from Eqs. (13) and (14)
\bea
\frac {dR} {dE_R} &=& \frac {\sigma_{\rm scalar}\rho_\chi} {4v_\oplus m_L
m_{\rm red}^2} F^2 (E_R) \l [ {\rm erf} \l ( \frac
{v_{\rm min} + v_\oplus} {v_0} \r ) \r. \nonumber \\
&&\l. - {\rm erf} \l ( \frac
{v_{\rm min} - v_\oplus} {v_0} \r ) \r ] \,\,\, .
\eea
The total local Dark Matter
density $\rho_\chi$ is taken to be 0.3 GeV/cm$^3$.
The above expression for differential rate
is for a monoatomic detector like Ge but it can be easily extended for
a diatomic detector like NaI as well.
                                                                                
The measured response of the detector by the scattering of Dark Matter 
off detector
nucleus is in fact a fraction of the actual recoil energy. Thus, the actual
recoil energy $E_R$ is quenched by a factor $qn_X$ (different for different
nucleus $X$) and we should express differential rate in Eq. (21) in terms of
$E = qn_XE_R$. For $^{76}$Ge, qn$_{\rm Ge}$ = 0.25 \cite{bot}, for 
$^{23}_{\rm Na}$, qn$_{\rm Na}$ = 0.3 \cite{brh}, for $^{127}$I,   
qn$_{\rm I}$ = 0.09 \cite{brh} and for $^{131}$Xe, 
qn$_{\rm Xe}$ = 0.8 \cite{bot}.   

Thus the differential rate in terms of the observed recoil energy 
$E$ for a monoatomic detector like Ge detector 
can be expressed as
\begin{equation}
\frac {\Delta R} {\Delta E} (E) =
\dis\int^{(E + \Delta E)/qn_{\rm Ge}}_{E/qn_{\rm Ge}}
\frac {dR_{\rm Ge}} {dE_R} (E_R) \frac {dE_R} {\Delta E}
\end{equation}
and for a diatomic detector like NaI, the above expression takes the form
\begin{eqnarray}
\frac {\Delta R} {\Delta E} (E) &=&
a_{\rm Na} \dis\int^{(E + \Delta E)/qn_{\rm Na}}_{E/qn_{\rm Na}}
\frac {dR_{\rm Na}} {dE_R} (E_R) \frac {dE_R} {\Delta E}  \nonumber \\
&+&a_{\rm I} \dis\int^{(E + \Delta E)/qn_{\rm I}}_{E/qn_{\rm I}}
\frac {dR_{\rm I}} {dE_R} (E_R) \frac {dE_R} {\Delta E}
\end{eqnarray}
where $a_{\rm Na}$ and  $a_{\rm I}$ are the mass fractions of Na and I
respectively in a NaI detector and are given by (see Table 2)
$$
a_{\rm Na} = \frac {m_{\rm Na}}
{m_{\rm Na} + m_{\rm I}} = 0.153 \,\,\,\,\,
a_{\rm I} = \frac {m_{\rm I}}
{m_{\rm Na} + m_{\rm I}} = 0.847
$$

The differential detection rates $\Delta R/\Delta E$ (/kg/day/keV)  
in case of LSH Dark Matter for different values of observed recoil 
energies are calculated using 
Eqs (10 - 22) for monoatomic detectors like Ge and Xe and using 
Eqs (10 - 23) for diatomic detector like NaI with $\Delta E = 1$ keV.
For calculation of the differential rates, we put $t=t_0$ in Eq. (17)
for all three types of detectors considered here. 
                                                                                
In Fig. 1 we plot LSH Dark Matter detection rate for  
$^{76}$Ge detector such as in GENIUS experiment 
at Gran Sasso. For demonstrative purpose we calculate the rates for 
three different LSH masses in $m_L < m_W$ regime,
namely $m_L =60$ GeV, 65 GeV and 70 GeV. 
For all the calculations the value of the coupling $|\lambda|$ is
fixed at 0.5. It appears 
from Fig. 1 that the differential rate decreases with the increase 
of LSH mass. 
This can be understood from Eq. (21) where the LSH mass $m_L$ 
appears at the denominator. We wish to make a comment here that 
while trying to plot this variation in logscale,   
it reveals 
that although the rates are more for lower mass of 
LSH Dark Matter and decreases with the increase of mass (as is 
evident form Fig. 1), the situation becomes reversed for high 
recoil energies as the rates show an oscillatory behaviour 
which becomes prominent at higher recoil energies.   
This phenomenon is due to the oscillatory nature of the 
Bessel function used in Eqs. (14) and (15). But at those high recoil energies
where the oscillatory nature becomes prominent, the yields are virtually nil.
Hence we do not plot them in linear scale instead.  

The variation of rates with detector recoil energies 
for diatomic NaI detector (used in DAMA experiment),
for three different LSH masses mentioned above, 
is shown in Fig. 2.
The coupling $|\lambda|$ is fixed at the same value of 0.5.

Similar results for Xenon ($^{131}$Xe) detector are plotted in Fig. 3.

One very positive signature of Dark Matter detection by direct detection 
method is the periodic annual variation of the detected Dark Matter. 
This periodicity arises due to the periodic motion of 
the earth around the sun by which the directionality of earth's motion 
changes continually throughout the year. Due to this, the amount of 
Dark Matter encountered
by the earth varies annually which can be understood from Eq. (17). This is 
the annual modulation of the detected Dark Matter and detecting this 
annual variation serves as a confirmatory test for Dark Matter detection.  
In order to see the possible annual variation of the detected LSH Dark 
Matter in three types of detectors discussed here, we have calculated the 
total events per day for a whole year for each of these three types of
detectors. Thus the value of $t$ in Eq. (17) 
varies from 1 to 365 while $t_0 = 153$ (2$^{\rm nd}$ June). 
For this purpose we have chosen a LSH mass of 65 GeV and the value of 
coupling to be $|\lambda| = 0.5$.
The results are plotted in 
Figs. 4, 5 and 6 for $^{76}$Ge, NaI and $^{131}$Xe detectors respectively.
The plots clearly show the sinusoidal behaviour of daily yield
over a year. It peaks in June and is minimum in December as expected.

For comparison, we calculate the rate and annual variation for one 
sample case for $m_L >> m_W$ regime and as discussed earlier this LSH cannot 
account for the total Dark Matter. Hence the local Dark Matter density
$\rho_\chi$ for this case is taken to be 0.003 GeV/cm$^3$ rather than
the usual total local Dark Matter density of  0.3 GeV/cm$^3$.  
The coupling $|\lambda|$ for 
this case is also kept at 0.5. Both the rate (Fig. 7) and the annual 
variation (Fig. 8) for this LSH clearly orders of magnitude lower 
than the cases considered in $m_L \la m_W$ regime (Figs 1 - 6).   
 
In summary, we investigate direct detection rates for a possible Dark Matter 
candidate in a Heavy Higgs model. The model 
contains an additional Higgs which neither couples with matter nor it has 
developed any VEV. This can be achieved through the choice of model 
parameters and discrete parity symmetry. Thus, the neutral part 
of the extra doublet becomes stable and can be a possible candidate for Cold 
Dark Matter. The LSH with mass around 60 - 70 GeV (in the regime 
$m_L \la m_W$) can explain the Dark Matter content of the universe. 
On the other hand if $m_L > m_W$, the LSH candidate may explain
only a fraction of Dark Matter.  
We calculated direct detection rates of those Dark Matter 
candidates in the context of three experiments namely $^{76}$Ge 
(as in GENIUS), DAMA and XENON. We have also shown the annual variation of 
the Dark Matter detection due to periodic motion of earth, 
for all the three experiments considered here. For comparison, 
the results for one case with $m_L = 300$ GeV ($>> m_W$)
is also shown.

\newpage
\begin{center}
{\bf Figure Captions} 
\end{center}

\noindent {\bf Fig. 1} Differential detection rates $\Delta R/\Delta E$ vs
recoil energy $E$ for three different values of LSH mass ($m_L$) 
namely 60 GeV, 65 GeV and 70 GeV. The topmost plot corresponds
to $m_L = 60$ GeV and the lowermost plot corresponds to
$m_L = 70$ GeV and the plot in between is for 
$m_L = 65$ GeV.
The coupling constant is kept fixed at 0.5 (see text).

\noindent {\bf Fig. 2} Same as Fig. 1 but for NaI. 

\noindent {\bf Fig. 3} Same as Fig. 1 but for $^{131}Xe$. 

\noindent {\bf Fig. 4} Annual variation of LSH Dark Matter direct detection
signal (events/kg/day vs each day in a year) for $^{76}$Ge detector
with $m_L = 65$ and $|\lambda| = 0.5$.

\noindent {\bf Fig. 5} Same as Fig. 4 but for NaI.

\noindent {\bf Fig. 6} Same as Fig. 4 but for $^{131}$Xe.
  
\noindent {\bf Fig. 7} Differential detection rates $\Delta R/\Delta E$ vs
recoil energy $E$ for LSH mass $m_L = 300$ GeV for $^{76}$Ge.

\noindent {\bf Fig. 8} Annual variation for LSH mass $m_L = 300$ at 
 $^{76}$Ge detector.

\end{document}